\documentclass[12pt,preprint]{aastex}

\usepackage{color}

\shorttitle{Preflare Nonthermal Sources}

\begin{document}

\title{Imaging Spectroscopy on Preflare Coronal Nonthermal Sources
Associated with the 2002 July 23 Flare}
\shortauthors{Asai et al.}

\author{
Ayumi Asai\altaffilmark{1}, 
Hiroshi Nakajima\altaffilmark{1}, 
Masumi Shimojo\altaffilmark{1}, 
Takaaki Yokoyama\altaffilmark{2}, 
Satoshi Masuda\altaffilmark{3}, 
and 
S\"{a}m Krucker\altaffilmark{4}}

\email{asai@nro.nao.ac.jp}
\altaffiltext{1}{
Nobeyama Solar Radio Observatory, National Astronomical Observatory of
Japan, Minamimaki, Minamisaku, Nagano, 384-1305, JAPAN}
\altaffiltext{2}{
Department of Earth and Planetary Science, University of Tokyo, 
Hongo, Bunkyo, Tokyo, 113-0033, JAPAN}
\altaffiltext{3}{
Solar-Terrestrial Environment Laboratory, Nagoya University, Chikusa, 
Nagoya, Aichi, 464-8601, JAPAN}
\altaffiltext{4}{
Space Sciences Laboratory, University of California, 
Berkeley, CA94720, USA}

\begin{abstract}
We present a detailed examination on the coronal nonthermal emissions
during the preflare phase of the X4.8 flare that occurred on 2002 July
23.
The microwave (17~GHz and 34~GHz) data obtained with Nobeyama
Radioheliograph, at Nobeyama Solar Radio Observatory and the hard
X-ray (HXR) data taken with {\it Reuven Ramaty High Energy Solar
Spectroscopic  Imager} obviously showed nonthermal sources that are
located above the flare loops during the preflare phase.
We performed imaging spectroscopic analyses on the nonthermal emission
sources both in microwaves and in HXRs, and confirmed that electrons
are accelerated from several tens of keV to more than 1~MeV even in
this phase.
If we assume the thin-target model for the HXR emission source, the
derived electron spectral indices ($\sim 4.7$) is the same value as
that from microwaves ($\sim 4.7$) within the observational
uncertainties, which implies that the distribution of the accelerated
electrons follows a single power-law.
The number density of the microwave-emitting electrons is, however, 
larger than that of the HXR-emitting electrons, unless we assume low
ambient plasma density of about $1.0 \times 10^9$~cm$^{-3}$ for the
HXR-emitting region.
If we adopt the thick-target model for the HXR emission source, on the
other hand, the electron spectral index ($\sim 6.7$) is much
different, while the gap of the number density of the accelerated
electrons is somewhat reduced.
\end{abstract}


\section{Introduction}
Nonthermal emissions from accelerated particles are often observed in
hard X-rays (HXRs), $\gamma$-rays, and microwaves at the beginning of a
solar flare.
Although these nonthermal emissions are undoubtedly associated with
intense energy release processes, the mechanisms to accelerate particles
are still unclear, and they have been one of the most important and the
most difficult problems in solar physics (see reviews by,
e.g., Aschwanden 2002). 
The HXR nonthermal emission is well explained with the bremsstrahlung
emission, which is emitted by the nonthermal electrons with energies $E
\gtrsim 20$~keV.
In the microwave range, on the other hand, the gyrosynchrotron emission
is the most promising nonthermal emission.
The microwave-emitting electrons have relatively higher energies, such as
sub-relativistic to relativistic energy.
Although the emission mechanisms and electron energies are totally
different, HXR and microwave emissions have shown a lot of similarities,
especially in the lightcurves (e.g., Kundu 1961).
The similarities have been thought to be evidences that
microwave-emitting electrons are accelerated with the same mechanism as
that for HXR-emitting electrons.
On the other hand, it has been also reported that an electron energy
spectral index derived from HXR emissions is often larger (softer) than
that derived from microwave emissions (e.g., Silva et al. 1997), and the
temporal behaviors of the spectral indices are totally different between
HXR and microwave.
These suggest a possibility that spectra of nonthermal electrons have a
bend, and that the high energy electrons that emit microwaves are
accelerated more efficiently than the HXR-emitting electrons.

To explain the gap of spectral indices, several models have been
suggested.
For example, \citet{Somov97} suggested that collapsing magnetic trap
works efficiently for particle acceleration of higher energy electrons.
Alternatively, \citet{Silv97} suggested that we should take into account
the transport mechanism of accelerated electrons, such as magnetic
trapping, since the nature of the emission mechanisms is different for
each emission range.
In HXRs, nonthermal emissions from footpoints of flare loops are
dominant, and these are described with ``thick-target'' bremsstrahlung
emission model.
However, the thick-target model only gives us a spectrum of the
injected electron flux from a HXR emission spectrum at rather lower
energy.
Therefore, the conversion from the injected electron flux to the
number of the nonthermal electrons is needed to compare it with the
number of the microwave-emitting electrons at higher energy, and it
requires another estimation such as the deflection time of
precipitating electrons.
If the deflection time has some dependences on electron energy, the
resulting electron spectral indices could be modulated.
Moreover, since the magnetic trapping works for a quite long time, and
sometimes lasts several tens of minutes, the microwave spectra suffer
modulations more, and it becomes more difficult to derive the
information of the electron acceleration in the later phase.

Recently, we examined the HXRs features of the 2002 July 23 flare, and
reported that the nonthermal energy even before the impulsive phase was
quite large \citep[Paper I]{Asa06}.
We refer to the time range when we can see the emission sources as
``preflare phase''.
We found sufficient emissions both in HXRs and in microwaves that can be
candidates for nonthermal emissions during the preflare phase.
The emission sources both in HXRs and in microwaves were located on the
flare loops, and the position almost corresponds to each other.
Examining the imaging spectroscopic features of the emission sources both
in HXRs and in microwaves and comparing the features with those of the
peak time are required to know particle acceleration mechanism in the
preflare phase.
In order to derive information on the nonthermal electrons in the preflare
phase, we examined in detail the features of the emission sources
spatially, temporally, and spectroscopically.
In this paper we report the results of the imaging spectroscopy on the
coronal emission sources observed in HXRs and in microwaves during the
preflare phase.
In \S 2 we describe the observational data, and we discuss the density
estimation in \S 3.
In \S 4 we present the detailed reports on the imaging spectroscopy of
the coronal emission sources both in microwaves and in HXRs.
In \S 5 we summarize our results and offer discussions.

\section{Observations}
\begin{figure}
\epsscale{.80}
\plotone{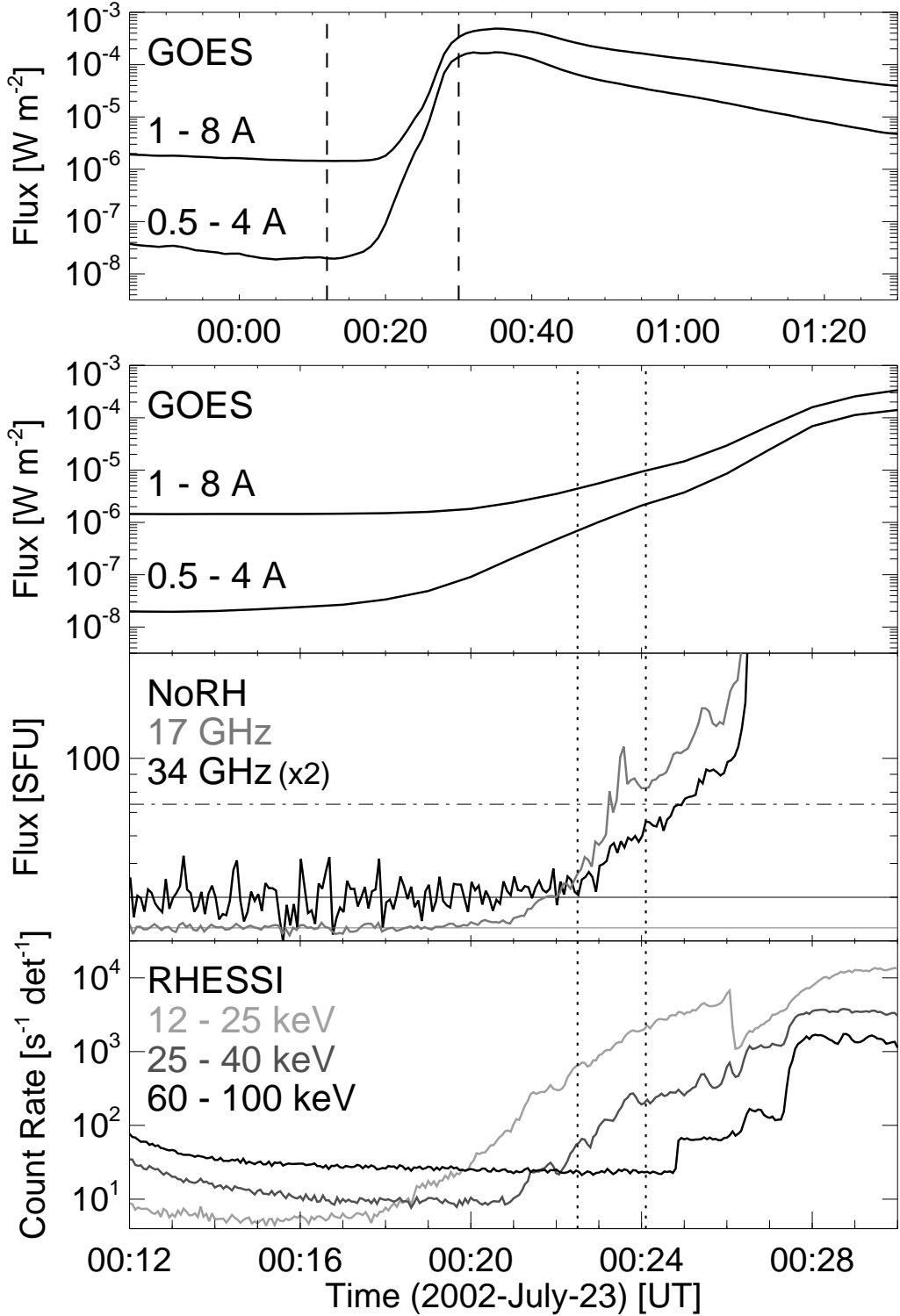}
\caption{
Lightcurves of the 2002 July 23 flare.
From top to bottom: 
soft X-ray flux in the {\it GOES} 1.0 -- 8.0~{\AA} channel; 
the same plot, but scaled up around the preflare phase that are indicated
with two {\it dashed} vertical lines in the top; 
radio flux observed at 17~GHz ({\it gray}) and 34~GHz ({\it black}) by
NoRH;
HXR count rate measured by {\it RHESSI} in 12 -- 25~keV ({\it light
gray}) , 25 --~40 keV ({\it dark gray}), and 60 -- 100~keV ({\it black}).
Two {\it dotted} vertical lines show the time range for the integration
of the {\it RHESSI} image synthesis.
The NoRH 34~GHz flux in the third panel is multiplied by 2, and
re-scaled so that the average flux, which was 73.8~SFU including the
quiet sun as shown with the dash-dot line, is to be 40.0~SFU in this
plot as shown with the solid line.
The average flux of the NoRH 17~GHz flux is about 32.7~SFU including
the quiet sun (as shown with the {\it gray} solid line).
\label{fig1}}
\end{figure}

The intense solar flare (X4.8 on the {\it GOES} scale) occurred in
NOAA Active Region 10039 (S12$^{\circ}$, E72$^{\circ}$) on 2002 July
23.
The {\it Reuven Ramaty High Energy Solar Spectroscopic Imager} ({\it
RHESSI}: Lin et al. 2002) showed us many spectacular features in HXR and
$\gamma$-ray wavelengths (e.g., Lin et al. 2003a).
This flare was also observed in microwaves with the Nobeyama
Radioheliograph (NoRH: Nakajima et al. 1994), as reported by
\citet{Whi03}.
The start time of this flare was recorded as 00:18~UT from the {\it
GOES} flux, and the impulsive phase that is defined by the intense
nonthermal emissions in HXRs and in microwaves started at about
00:27:30~UT.
Figure~\ref{fig1} shows time profiles of the flare in SXRs, microwaves,
and HXRs.
The top panel shows the lightcurves in SXRs taken with the {\it GOES} 1.0
-- 8.0~{\AA} (top) and 0.5 -- 4.0~{\AA} (bottom) channels.
The bottom three panels of Figure~\ref{fig1} show the expanded
lightcurves of the rise phase of the flare (from 00:12~UT to 00:30~UT),
which corresponds to the time between the two {\it dashed} lines in the
top panel.
The top of the three panels shows the lightcurves of the {\it GOES} 1.0
-- 8.0~{\AA} and 0.5 -- 4.0~{\AA} channels, the middle panel shows the
ones at NoRH 17~GHz and 34~GHz, and the bottom one shows the {\it
RHESSI} time profiles in three energy ranges of 12 -- 25, 25 -- 40, and
60 -- 100~keV.
In this paper we focus on the nonthermal emissions in HXRs and in
microwaves of the preflare phase, from 00:22:30 to 00:24:06~UT, on 2002
July 23, which corresponds to the time between the two {\it dotted}
lines in the bottom three panels.
This time range corresponds to the Phase II and III in Paper I, and we
can identify the first nonthermal emissions and the onset of the faint
ejection in the extreme ultraviolet images during this phase.

\begin{figure}
\plotone{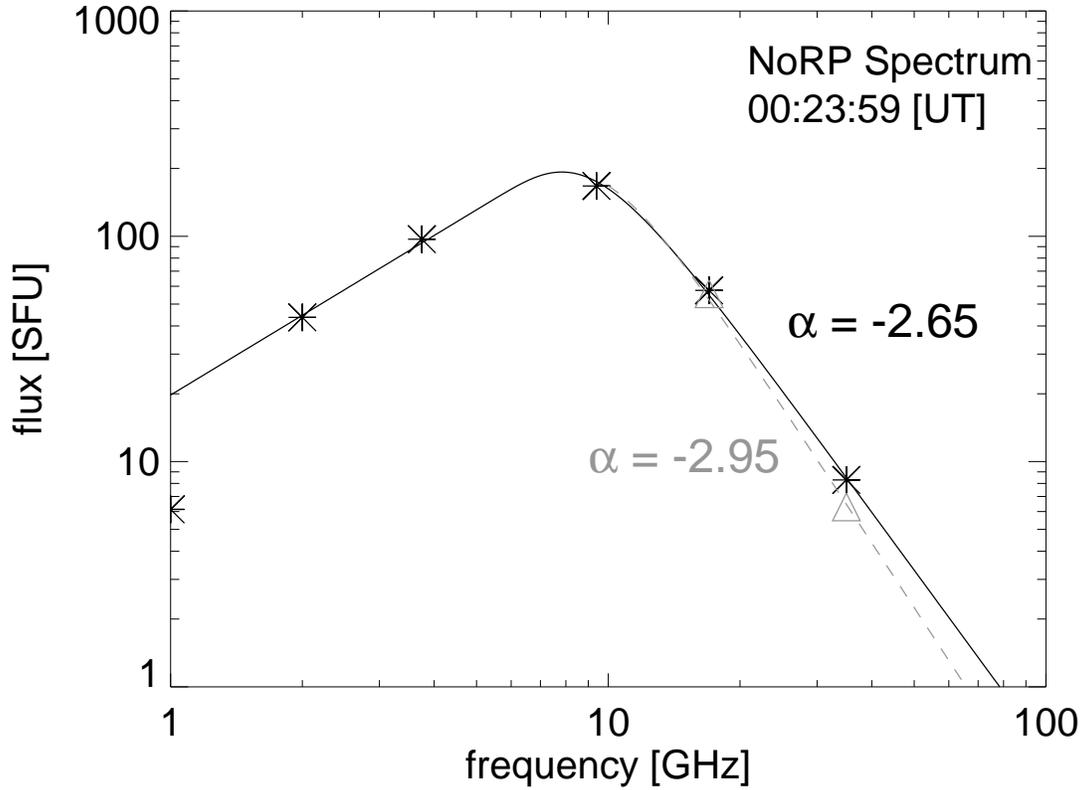}
\caption{
Spectrum taken at 00:23:59~UT by NoRP.
The fitted parameters are as follows:
the index $\alpha$ for the optically-thick part is 1.17,
that for the optically-thin part is -2.65,
the turn-over frequency is 9.4~GHz, and 
the peak flux is 276~SFU.
Subtracting the thermal microwave emission, which is about 2~SFU, we
re-plotted the spectrum with the index $\alpha$ for the optically-thin
part of -2.95, as shown with the {\it gray} dashed line.
\label{fig2}}
\end{figure}

Figure~\ref{fig2} presents a microwave spectrum taken at 00:23:59~UT by
the Nobeyama Radio Polarimeters (NoRP; Torii et al. 1979; Shibasaki et
al. 1979; Nakajima et al. 1985).
NoRP measure the total fluxes of the flare at 1, 2, 3.75, 9.4, 17, 35
and 80~GHz with a temporal resolution of 0.1~second.
We fitted the spectrum by using the NoRP data of 2, 3.75, 9.4, 17 and
35~GHz, and obtained the spectral index $\alpha$ ($F_{\nu} \propto
\nu^{\alpha}$; $F_{\nu}$ is the flux density at frequency $\nu$) of
$-2.65$ and the turn-over frequency of about 9~GHz.
As is seen in Figure~\ref{fig1} and will be mentioned more below, during
the time range on which we focus, the thermal emission rapidly increases
and the thermal component also emits microwaves.
We estimated the microwave thermal emission observed with NoRP at 17
and 35~GHz by using the {\it GOES} data and it is about 2~SFU.
Since this somewhat hardens the microwave spectrum, we modified the
NoRP spectrum as shown with the dashed line in Figure 2, and the
fitted spectral index $\alpha$ in the optically-thin part is a little
steepened to be about -2.95.
We re-plotted the fitting result on Figure~\ref{fig2} with the {\it
gray} dashed line.

On the other hand, NoRH observes the Sun at two frequencies, 17 and
34~GHz during Japanese daytime (normally from 22:45 to 06:30~UT) with a
temporal resolution of 1~second.
We can derive a spectral index $\alpha$ by using the flux ratio.
The derived index $\alpha$ is for the optically-thin gyrosynchrotron
emission, since both the 17 and 34~GHz emissions are in the
optically-thin part.
The spatial resolutions (FWHMs of the synthesized beam) of NoRH data are
$14^{\prime\prime}$ for 17~GHz and $7^{\prime\prime}$ for 34~GHz.

\citet{Lin03b} and \citet{Hol03} already reported on the HXR emission
in the preflare phase, although their works were based on the {\it
spatially integrated} emission.
We also synthesized the HXR images obtained with {\it RHESSI} by using
grids 3 -- 8 and Clean method, which gives the spatial resolution (FWHM)
of about $10^{\prime\prime}$.
EUV images of the flare were obtained with the {\it Transition Region
and Coronal Explorer} ({\it TRACE}; Handy et al. 1999; Schrijver et
al. 1999).
We used 195~{\AA} images, in which the Fe~{\sc xii} line formed at
$\sim$1~MK is normally dominant.
The pixel size of the CCD is $1^{\prime\prime}_{\,\cdot}0$, and 
the temporal resolution is about 9~second.

\begin{figure}
\plotone{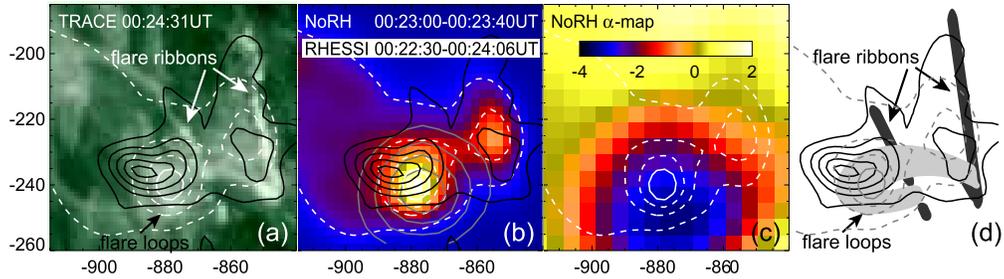}
\caption{
Images for the preflare phase of the flare.
Solar north is up, and west is to the right.
(a) shows the EUV image taken with {\it TRACE} 195~{\AA} at 00:24:31~UT.
(b) (c) show the microwave image of NoRH 34 GHz and the map of NoRH
$\alpha$ index.
(d) shows the positions of the flare ribbons ({\it dark gray} regions)
and the flare loops ({\it light gray}).
On each panel we overlaid the contour images of NoRH 34~GHz with the
{\it white} dashed line (with the {\it gray} dashed line for d).
The NoRH 17~GHz contour image is further overlaid on the panel (b) with
the {\it gray} solid line.
The levels of the NoRH contours are 20, 40, 60, and 80~\% of the peak
intensity.
We also showed the core region of the microwave loop-top emission source
with the {\it white} solid line that is the 80~\% contour line.
We also overlaid the contour images of the {\it RHESSI} 30 -- 40~keV
intensity  and {\it black} solid line on (a), (b), and (d).
The {\it RHESSI} contours are 20, 40, 60, 80, and 95~\% of the peak
intensity.
\label{fig3}}
\end{figure}

Here, we summarize the spatial features of the emission sources.
Figure~\ref{fig3} shows images of the phase in EUV, HXR, and microwave.
The panel (a) is the {\it TRACE} 195~{\AA} image taken at 00:24:31~UT.
We can see a large two-ribbon structure that brightens from 00:20~UT.
We can also see a diffuse loop-like structure that is identified as
Fe~{\sc xxiv} emission from 20~MK plasma, as is often observed in {\it
TRACE} 195~{\AA} images during the impulsive phase of a flare.
The diffuse loop seems to connect the two-ribbon structure as shown in
the cartoon of the panel (d).
We overlaid a microwave contour images of NoRH 34~GHz on each panel with
the {\it white} dashed line (while it is the {\it gray} dashed line in
the panel d).
We also overlaid the NoRH 17~GHz contour image on the panel (b) with
the {\it gray} solid line.
The levels of the contours are 20, 40, 60, 80~\% of the maximum
intensity, and only the highest level (80~\%) of the 34~GHz contours is
shown with the solid line in each panel.
In addition, on panels (a), (b), and (d), we overlaid an HXR contour
image observed with {\it RHESSI} in 30 -- 45~keV with the {\it black}
solid line.
The levels of the contours are 40, 60, 80, 95~\% of the maximum intensity.
We recognize a microwave and an HXR emission sources appear above the
flare ribbons \citep{Lin03b}.
The position of the microwave emission source is slightly lower than
that of the HXR emission source ($\sim 10^{\prime\prime}$),
which corresponds to the top of the post flare loops that become
visible in the later phase in the {\it TRACE} images (see also Paper
I).
The footpoint emission sources are much weaker and unclearer than the
loop-top sources.
We can see compact emission sources on the western flare ribbon both in
HXRs and in microwaves, and we can marginally see the extension of the
HXR contour line that outlines the eastern flare ribbon.
The microwave contour image with the lowest level also outlines a large 
loop-like bright region in the NoRH image.
This remains visible for several hours before this flare started
\citep{Whi03}.

\section{Density Estimation}
Here, we present the density estimation of the ambient corona where the
nonthermal emission sources appear.
Before the flare in question started (from 22:00~UT on 2002 July 22),
the {\it GOES} SXR fluxes increased, and this probably comes from the
emission source which is visible as the large loop-like bright region in
the microwave \citep{Whi03}.
This feature seen in the before-the-flare phase could be related to a
small flare that occurred at 22:00~UT in the same active region,
although we could not confirm this due to a lack of image data for the
small event.
We estimated the temperature and the emission measure of this emission
source, by using the ratios of the two of {\it GOES} channels, and by
defining the emission from the background corona as the average emission
between 20:30 and 21:10~UT on 2002 July 22.
Then, the temperature and the emission measure are estimated to be about
5.7~MK and $8.0 \times 10^{48}$~cm$^{-3}$ at 23:30~UT, respectively.
As we already reported in Paper I, we could estimate the microwave
fluxes in 17 and 34~GHz from the temperature and the emission measure,
by assuming that the free-free emission is dominant for the microwave
source, and confirmed that they were almost the same as the observed
values.

After 00:18~UT, on the other hand, the {\it GOES} temperature rapidly
increases and is higher than 10~MK after 00:21~UT.
This means that hot thermal plasma ($T > 10$~MK) is also generated as
well as the nonthermal emissions seen in HXRs and in microwaves.
The accelerated nonthermal electrons probably traveled in this hot
plasma, and therefore, the density estimation of the hot plasma is
required for further discussions.
To derive the temperature and the emission measure of the hot component,
we re-defined the emission from the background corona as the emission
between 23:30 and 00:10~UT.
They are estimated about 15~MK and $2.4 \times 10^{48}$ cm$^{-3}$ at
00:23~UT.
To derive the ambient plasma density, we assume the source volume.
The size of the HXR source is about $27^{\prime\prime}$ (in east-west
direction) $\times$ $20^{\prime\prime}$ (in north-south direction).
Hence, assuming the line-of-sight depth of the emission source as
$1.5 \times 10^{9}$ cm, the volume and the density are about $4.2 \times
10^{27}$~cm$^{3}$ and $2.4 \times 10^{10}$~cm$^{-3}$.

\section{Imaging Spectroscopy}

\subsection{Microwave Emission Source}
First, we examine the spectral features of the microwave emission
source by using the NoRH data.
For the analyses, we synthesized the images both at 17 and 34~GHz from
00:23:00 to 00:23:40~UT (40 second duration) every 1~second, and
integrated them to make one image at each frequency.
Then, we conformed the beam size of the 34~GHz image to that of the
17~GHz image to take ratio of the intensities, by convolving each image
by the beam of the other image.
Roughly speaking, this degrades the spatial resolution of the 34~GHz
image, which is comparable to that of the 17~GHz image ($\sim
14^{\prime\prime}$).
Finally, we can derive a spectral index $\alpha$ by using the flux
ratio, i.e., $\log(F_{34 \mathrm{GHz}}/F_{17 \mathrm{GHz}}) (\log(34
\mathrm{GHz}/17 \mathrm{GHz}))^{-1}$.

Figure~\ref{fig3}c shows the map of the index $\alpha$ (we call it
``$\alpha$-map'') obtained with NoRH, overlaid with the NoRH 34~GHz
contour image with the {\it white} line.
The core emission area is selected to be larger than 80~\% of the maximum
intensity of 34~GHz image, which corresponds to the inside region of the
innermost ({\it solid}) contour line.
We measure the spectral index $\alpha$ of the emission source, and
found that it is about $-3.0$, which is consistent with that derived
from the NoRP spectrum.
The center of the core emission region is ($-879$, $-241$) arcsec
heliocentric.
The area is about $5.1 \times 10^{17}$~cm$^2$ on the solar surface, and
this corresponds to the solid angle $\omega$ of about $2.3 \times
10^{-9}$~str.
The total fluxes of the region at 17~GHz and at 34~GHz are 
$F_{17} \sim 4.5 \times 10^{-19}$ and $F_{34} \sim 5.1 \times
10^{-20}$~erg~s$^{-1}$~cm$^{-2}$~Hz$^{-1}$ (4.5 and 0.51~SFU),
respectively. 
The maximum brightness temperature observed at the NoRH 17~GHz is $3.1
\times 10^6$~K and that at the 34 GHz $1.6 \times 10^5$ K.

Here, we mention on the error to estimate the spectral index
$\alpha$.
The error is mainly caused by the relative displacement between the
images at 17~GHz and those at the 34~GHz due to the NoRH image
syntheses.
The NoRH image syntheses hold an uncertainty of the positioning of about
$5^{\prime\prime}$, and in this case, the error on $\alpha$ is about
$\pm$0.15.

If we assume that the accelerated electrons follow a power-law
distribution $dn_{\mu}(E)/dE = K E^{-\delta_{\mu}}$
electrons~cm$^{-3}$~keV$^{-1}$ ($K$ is a constant), the optically-thin
gyrosynchrotron emission also follow a power-law distribution with the
spectral index $\alpha$.
There have been several studies to derive the relation between
$\delta_{\mu}$ and $\alpha$, and we adopt the approximation derived by
\citet{Dulk85} here.
For the calculations, the number density of total electrons with higher
energy than the lower-energy cutoff $E_c$ ($E > E_c$), that is,
$N_{E_c}$ ($=\int_{E_c}^{\infty} dn_{\mu}(E^{\prime})/dE^{\prime}
dE^{\prime}$), is often used instead of $n_{\mu}(E)$, and \citet{Dulk85}
assumed that $E_c = 10$~keV.
The constant $K$ is also related to $N_{E_c}$ with the relation $K =
(\delta - 1) E_c^{\delta -1} N_{E_c}$.
\citet{Dulk85} showed $F_{\nu} \propto \nu^{1.22-0.90\delta_{\mu}}$, and
therefore, we derive $\delta_{\mu} \sim (1.22 - \alpha) / 0.90 \sim 4.7$
in the present case.
From these, we can also estimate $N_{\rm 10 keV}$ as follows:
\begin{equation}
N_{\rm 10 keV} \sim 3.4 \times 10^{42} (\sin \theta)^{-2.6}
 \frac{F_{\nu} \nu^{3.0}}{\omega l B^{4.0}}
 \sim 3.1 \times 10^{18} (\sin \theta)^{-2.6} B^{-4.0},
\end{equation}
where $\theta$ is the angle between the magnetic field $B$ and the line
of sight, and $l$ is the line-of-sight depth of the emission source and
is assumed here to be comparable to the width of the bundle of the EUV
flare loops seen in the {\it TRACE} images, and about
20$^{\prime\prime}$ ($\sim 1.5 \times 10^{9}$~cm).
Then, the constant $K$ is rewritten as $6.1 \times 10^{22}(\sin
\theta)^{-2.6} B^{-4.0}$.

The gyrosynchrotron emission strongly depends on the magnetic field
strength in the corona, which is very difficult to correctly measure.
We use the microwave emission of the impulsive phase of the flare
observed with NoRH to measure it.
Fortunately, we observed the optically-thin thermal emission
\citep{Whi03,Asa06} in the before-the-flare phase (which corresponds to
the Phase I in Paper I), which shows the circular polarization depending
on the magnetic field strength \citep{Dulk85}.
The degree of the circular polarization measured for the NoRH 17 GHz
emission source is about 1.6~\%, which means $B_{0} \cos \theta \sim
48$~gauss.
The estimation of $B_{0}$ depends on $\theta$, and becomes $\sim$ 190
and 68 gauss for the cases of $\theta$ = 75 and 45$^{\circ}$,
respectively.
For the further estimations in this paper, we used the $B_{0} \cos
\theta$ measured in the before-the-flare phase as that in the preflare
phase, although the magnetic field during the phase could be much
different.
If we assume that $B_{0} = 150$ gauss and $\theta = 71^{\circ}_{\,\cdot}0$, 
we found that 
\begin{equation}
\frac{dn_{\mu}(E)}{dE} = 1.2 \times 10^{14} E^{-4.7}
\end{equation}
electrons~cm$^{-3}$~keV$^{-1}$.
In this paper we assume that $E_{c}$ is 20~keV, and then, we got $N_{\rm
20keV} \sim 4.8 \times 10^{8}$ electrons~cm$^{-3}$.
From the spectra of the HXR total flux, \citet{Hol03} also derived
$E_{c} \sim$ 20~keV in this phase of the flare.
We reported other cases with different magnetic field strengths $B$ =
100, 150, and 200~gauss, and summarized the results in Table 1.
We can easily confirm that the $N_{\rm E_c}$ strongly depends on the
magnetic field strength.

To generate the gyrosynchrotron emission observed in microwaves, such as
at 17~GHz, with these magnetic field strengths (100 -- 200 gauss),
nonthermal electrons have to be accelerated to more than 1~MeV
\citep{Kosu88,Bas99}.
Therefore, we conclude that electrons are effectively accelerated to
such high energy, even in the preflare phase of the flare.
\citet{Whi03} reported that the energy spectral index derived from NoRP,
$\delta_{\mu}$ is about 2.7 -- 1.8 during the impulsive phase of the
flare.
However, these values could be underestimated due to the too high
turn-over frequency which is reaching up to 30~GHz.
On the other hand, the spectral index derived from the 40 -- 400 keV
HXR, $\delta_{H}$ is about 4.5 during the peak time, which is almost the
same as $\delta_{\mu}$ we obtained ($\sim 4.7$).
This implies that we observed the nonthermal electrons, which become the
main component in the impulsive phase, even in the preflare phase.

\subsection{HXR Emission Source}
Second, we investigate the spectral index of the HXR emission source
during the preflare by using the {\it RHESSI} data.
In HXR range, we often observe nonthermal emission that follow a
power-law distribution with the energy spectral index $\gamma$.
The intensity $I({\epsilon})$ photons~s$^{-1}$~cm$^{-1}$~keV$^{-1}$ is
written as $I(\epsilon) = a \epsilon^{-\gamma}$, where $\epsilon$ is
energy of the photon and $a$ is constant.

\begin{figure}
\plotone{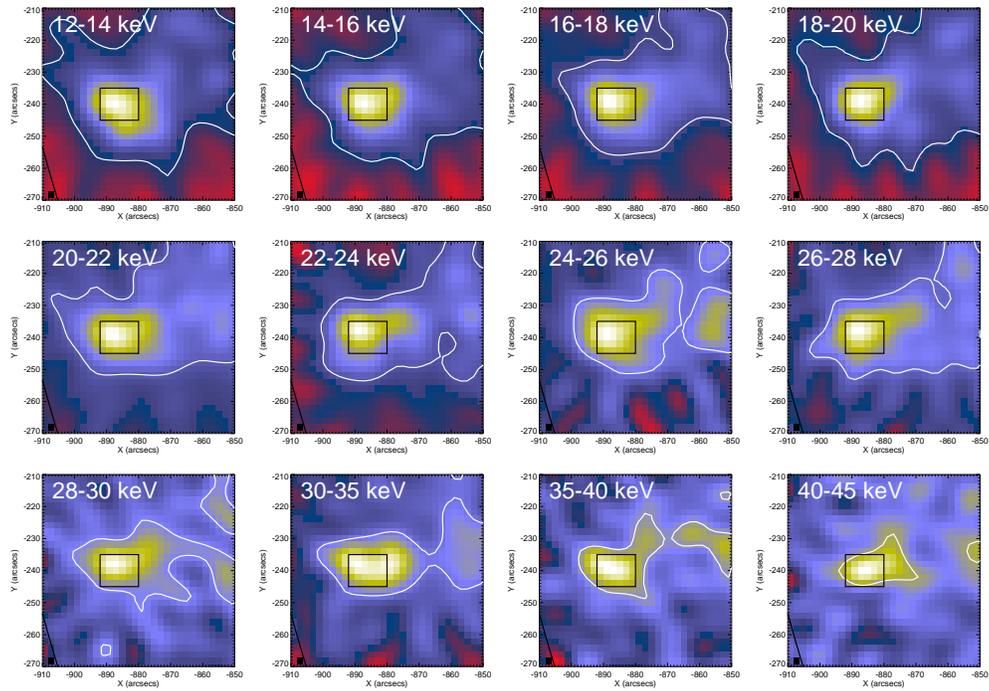}
\caption{
HXR images in each energy bin from 12 to 45~keV.
The rectangles with the {\it black} lines show the core emission region.
The {\it white} contour lines show 3$\sigma$ level of each HXR image.
\label{fig4}}
\end{figure}

\begin{figure}
\plotone{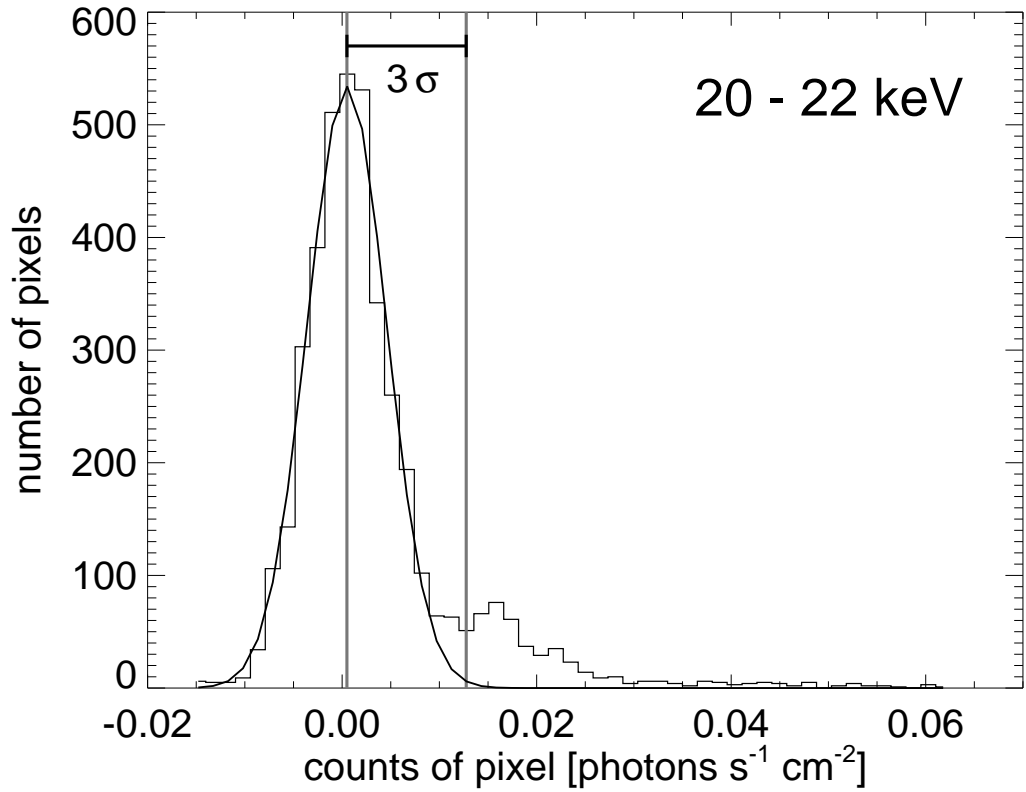}
\caption{
Distribution histogram of the {\it RHESSI} image in 20 -- 22~keV band.
We overlaid Gaussian function with the {\it solid} line.
The 3$\sigma$ level of the Gaussian function is also shown with the
vertical line.
\label{fig5}}
\end{figure}

Here, we explain the way of the imaging spectroscopy of the HXR data.
This method was originally suggested by \citet{Mita05}.
We synthesized the HXR images in 2 keV-bin from 10 to 30~keV, and in 5
keV-bin from 30 to 50~keV.
We integrated over 96 seconds, from 00:22:30 to 00:24:06~UT to
synthesize the images in each energy bin.
Figure~\ref{fig4} shows the image in each energy band.
We can see that the loop top emission sources are dominant in each
energy band.
We can also see faint components elongated in the north-south direction
in some images that probably outline the flare ribbons.
Then, we draw a distribution histogram on the photon counts of an image
in each energy band.
We present an example in Figure~\ref{fig5}, which is the distribution
histogram of the image in 20 -- 22~keV band.
We fit the histogram with a Gaussian function, and determined the noise
level of the image as 3$\sigma$ of the Gaussian function as shown in
Figure~\ref{fig5}.
We selected the core emission region of the HXR images as shown with the
box in Figure~\ref{fig4}.
All of the pixels within the core emission area are larger than the
3$\sigma$ level of the images from 10 -- 12~keV band to 40 -- 45~keV
band.
The center of the core region is ($-886$, $-239$) arcsec heliocentric,
where is slightly higher in vertical direction than the microwave core
region.
The image of 45 -- 50~keV is noisy and there are few pixels that exceeds
the noise levels, so that we cannot distinguish the coronal emission
source in the image.
Therefore, we did not use the image for further spectral analyses.

\begin{figure}
\plotone{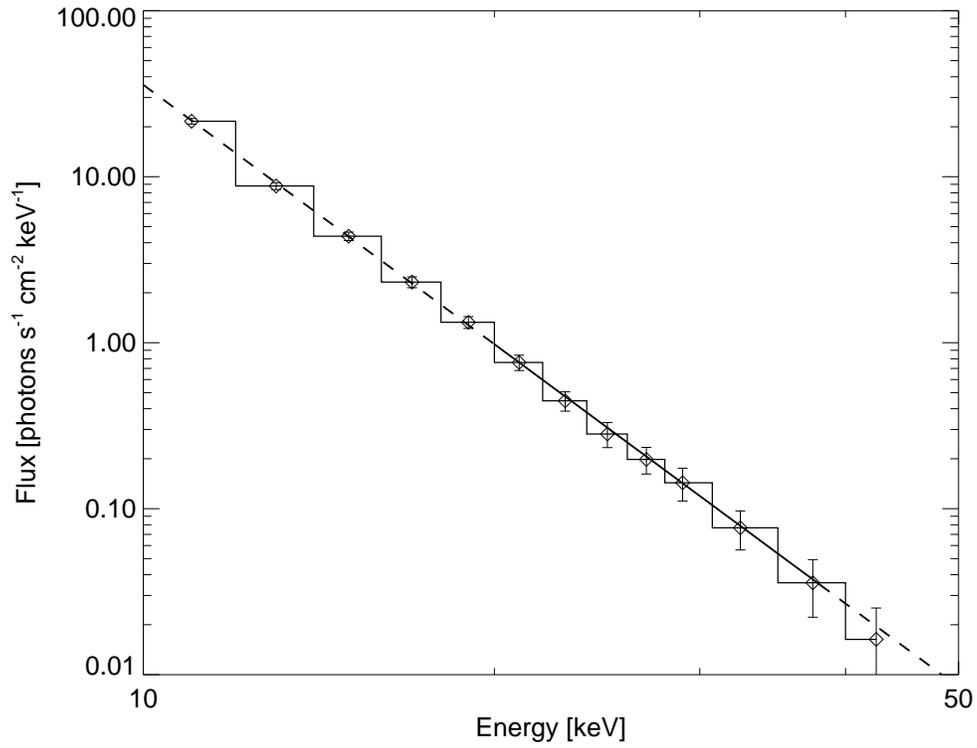}
\caption{
Photon spectrum of HXR emission region overlaid with the fitting plot.
The time interval is from 00:22:30 to 00:24:06~UT (96~sec).
The counts for the fit are integrated over the core emission region
shown in Figure~\ref{fig4}.
The error bars show the quadrature of the 3$\sigma$ level of each energy
bin and the 3$\sigma$ level of the photon noise.
\label{fig6}}
\end{figure}

The intensity of the core region clearly shows the power-law
distribution as shown in Figure~\ref{fig6}.
We fitted the spectrum from 20 to 40~keV range with a single power-law.
As a result, we got $I(\epsilon) = 3.2 \times 10^{6} \epsilon^{-\gamma}$, 
and $\gamma$ is $5.2 \pm0.1$.
We also showed the fitting result in Figure~\ref{fig6}.
The error bars on the plot show the 3$\sigma$ levels including the
photon noise of the signal.
This value ($\sim 5.2$) roughly corresponds to the result that was
derived from the spectroscopy of the total flux (no imaging) for the
preflare phase of this flare \citep{Hol03}, while their other results,
i.e., the break of the spectra around 30~keV and the softening at the
higher energy range, are unclear in our observations.
It may be because our result is based on the imaging spectroscopy from
10 to 40~keV, and the softening in the higher energy range does not
appear clearly in the energy range.
Alternatively, the single power-law fit that we derived may be just a
weighted average of the spectral indices within this energy range, and
the break and the steeper spectra at higher energy range may be missed.
In addition, although the thermal component is also unclear in our
result, it is consistent with the results by \citet{Lin03b} and
\citet{Hol03} who reported that they could fit the spectra by the double
power-law without any thermal components before 00:26~UT.

We could adopt the thin-target bremsstrahlung model to convert $\gamma$
into an electron power-law index $\delta_{H}$, since the HXR source is
located on the loop-top, where is almost the same location as the
nonthermal emission source seen in microwaves.
Assuming the thin-target model for the HXR emission suggested by
\citet{Hud78}, the spectrum of the accelerated electrons $dn_{H}(E)
V/dE$ electrons keV$^{-1}$ ($V$ is the volume of the emission source) is
written as follows:
\begin{equation}
\frac{dn_{H}(E) V}{dE} = 1.05 \times 10^{42} \frac{a}{n_{i}} 
 \frac{\gamma-1}{B(\gamma-1,\frac{1}{2})} E^{-\gamma + 0.5},
\end{equation}
where $E$ is energy of the electron, $n_{i}$ is the number density of
ambient target plasma, and $B(p,q)$ is the Beta-function.
Moreover, this model suggests that $dn_{H}(E)/dE = K^{\prime}
E^{-\delta_{H}}$ ($K^{\prime}$ is a constant), and therefore, we found
$\delta_{H} = \gamma - 0.5 \sim 4.7$.
Then, the electron spectrum is rewritten as 
$dn_{H}(E)/dE = 2.7 \times 10^{49} n_{i}^{-1} V^{-1} E^{-4.7}$.
The electron spectral index derived from the {\it RHESSI} HXR data
$\delta_{H}$ is almost the same as that from the NoRH microwave data
$\delta_{\mu}$.
This implies that the distribution of the accelerated electrons follows
a single power-law.

If the thin-target model is a reasonable assumption to the HXR emission
source, we can further estimate the number density
$dn_{H}/dE$~electrons~cm$^{-3}$~keV$^{-1}$ of the HXR-emitting electron
in the core region.
Assuming again that the width in the line-of-sight direction is
comparable to the EUV flare loops seen in the {\it TRACE} images ($\sim
1.5 \times 10^{9}$~cm), the volume $V$ is estimated as $9.6 \times
10^{26}$ cm$^3$.
As we mentioned in \S 2, the density of the ambient plasma $n_{i}$ is
roughly estimated from the {\it GOES} emission, and is about $2.4 \times
10^{10}$~cm$^{-3}$.
Therefore, we found that
\begin{equation}
\frac{dn_{H}(E)}{dE} = 1.2 \times 10^{12} E^{-4.7}
\end{equation}
electrons~cm$^{-3}$~keV$^{-1}$.
Then, we also estimate the number density $N_{E_c}$ of nonthermal
electrons with $E > E_{c}$.
Since $N_{E_c} = \int_{E_c}^{\infty} dn_{H}(E^{\prime})/dE^{\prime}
dE^{\prime}$, the result strongly depends on the lower-energy cutoff
$E_c$.
Here, we assume the case of $E_{c} =$ 20~keV, and we found that $N_{\rm
20keV} = 5.1 \times 10^{6}$ electrons~cm$^{-3}$.
These results on the accelerated electrons are about two orders of
magnitude less than those from the microwave emission (c.f. eq. 2).
Therefore, it may be better to consider that the emissions showed the
same spectral indices by an accidental coincidence.

Then, we consider the thick-target model to explain the HXR emission
source.
The accelerated nonthermal electrons seem to be effectively trapped in
the corona in this phase, since we cannot clearly see footpoint sources
but loop-top sources (see Fig.~3).
Therefore, we suggest that the trapped nonthermal electrons are
exhausted in the corona with the thick-target model, and apply the
trap and precipitation model or the thick-thin target model
\citep{Mel76,Asch98,Whea95,Met99}.
The thick-thin target model suggests that a break, which increases
with time from about 15~keV to 40~keV or more, in a HXR spectrum, and
thick-(thin-) target model should be adopted below (above) the break
energy.
\citet{Hol03} reported that there is a break of the total photon
spectra at around 30~keV, which probably shows the transitions from
the thick-target to the thin-target, while it is unclear in our
imaging spectroscopic study.
Following the eq. 1 of \citet{Met99}, the break energy is estimated 
$E_{t} = 20 \times (n_{i} l_{e}/10^{20})^{0.5}(0.7/\cos \alpha_{0})^{0.5}$~keV,
where $l_{e}$ is the length that an accelerated electron travels, and
$\alpha_{0}$ is the average pitch angle of the trapped electrons.
We roughly estimate that $l_{e}$ is twice as long as the height of
the flare loops seen in the {\it TRACE} images (i.e., we assume that the
the X-point is located twice as high as the loop height, and that the
accelerated electrons are thermalized before they bombard at the
footpoints), and is about $80^{\prime\prime}$ ($\sim 5.8 \times
10^{9}$~cm).
Therefore, we get $E_{t} \sim 24 \times (0.7/\cos
\alpha_{0})^{0.5}$~keV, and it is, for example, about 34~keV for
$\alpha_{0} \sim 70^{\circ}$.

In the case of the thick-target model, the power-law distribution of the
injected electron flux $F(E)$ electrons~s$^{-1}$~keV$^{-1}$ is written
as \citep{Hud78}:
\begin{equation}
F(E) = \frac{d^{2}n_{H}(E) V}{dE dt} = 3.28 \times 10^{33} a \gamma^{2}
 (\gamma-1)^{2} B(\gamma-\frac{1}{2},\frac{3}{2}) E^{-\gamma - 1.0}.
\end{equation}
In the present case we get
\begin{equation}
F(E) = \frac{d^2n_{H}(E) V}{dE dt} = 7.0 \times 10^{41} E^{-6.2},
\end{equation}
and the spectral index $\delta_{H}$ is about 6.2.
Moreover, we get $dn_{H}(E)/dE \sim F(E) \tau V^{-1} = 5.8 \times
10^{14} E^{-6.7}$ electrons~cm$^{-3}$~keV$^{-1}$, assuming that $\tau$
is the electron traveling time and $\tau \sim l v_{e} = l
\sqrt{m_{e}/(2E)}$, where $v_{e}$ and $m_{e}$ are electron velocity
and mass, and the traveling length $l$ (which is assumed to be $1.5
\times 10^{9}$ cm).
In this case the total number
of the accelerated electrons ($N_{\rm 20keV} = 4.0 \times 10^{6}$
electrons~cm$^{-3}$) is somewhat closer to that derived from the
microwave than those obtained in the thin-target case, while the
spectral index $\delta_{H}$ ($\sim 6.7$) is much different from those
for the microwave-emitting electrons.


In the impulsive phase of the flare, the {\it RHESSI} HXR spectra are
well fitted with thermal plus double power-law distributions
\citep{Hol03,Whi03}.
Especially, the coronal source no longer shows the nonthermal features,
and is responsible for the thermal component in the spectrum
\citep{Ems03}.
The power-law components, which mainly come from the footpoint sources,
show a quite harder spectrum with $\gamma \sim 3$ than that for the
preflare phase, and the corresponding electron energy spectral index is
about $\delta \sim 4.5$, assuming the thick-target model for the HXR
emission.
As we mentioned above, this is very close to the spectral index derived
from the microwave emission in the preflare phase ($\sim 4.7$).

\section{Summary and Discussions}
We performed imaging spectroscopic analyses on the emission sources
observed both in NoRH microwaves and in {\it RHESSI} HXRs.
Both the emission sources are located above the post flare loops in the
corona, although the HXR emission source is located slightly higher, and
they clearly show nonthermal features.
Based on the gyrosynchrotron theory \citep{Kosu88,Bas99}, nonthermal
electrons have to be accelerated to higher than 1~MeV even in this phase.

If we assume the thin-target model for the HXR emission source, the
electron spectral index $\delta_{H}$ of about 4.7 showed the same value
as that from microwaves $\delta_{\mu}$ ($\sim 4.7$) within the
observational uncertainties.
This result implies that the distribution of the accelerated electrons
follows a single power-law.
The number density of the nonthermal electrons that emit the microwaves
is, however, much larger than that of the HXR-emitting electrons.

If the real magnetic field strength is higher than our estimation,
the gap between these indices reduces.
For example, if we assume very strong magnetic field of about 300~gauss
($\theta = 81^{\circ}$), then, $N_{20keV} = 2.7 \times 10^{7}$, which is
somewhat close to the result derived from the thin-target model.
However, as \citet{Whi03} reported, such high magnetic field is not
expected in the impulsive phase of the flare from the gyrosynchrotron
theory \citep{Dulk82}, and they also commented that magnetic field
strength is probably no more than 200~gauss.
Although we cannot directly adopt the same discussion to the preflare
phase, magnetic field strength of about 300~gauss is somewhat unusual.
Furthermore, the {\it TRACE} 195 {\AA} images do not show that the flare
loops stand in a direction exactly perpendicular to the line-of-sight.
Therefore, the larger $\theta$ (such as 80$^{\circ}$ or more), and
therefore, the strong magnetic field cannot be expected.

We may overestimate the ambient plasma density.
If we assume the number density of about $n_{i} = 1 \times
10^{9}$~cm$^{-3}$, we found that $N_{\rm 20keV} = 1.2 \times
10^{8}$~electrons cm$^{-3}$ for the thin-target models.
These are comparable to the results from $n_{\mu}$ with the case of $B =
200$ gauss (see Table 1).
On the other hand, we can see faint flare loops and two-ribbon structure
in {\it TRACE} images, which implies that hot plasma with the
temperature of about 20~MK fills the whole arcade region, even in the
preflare phase.
Then, we can estimate the maximum size of the hot plasma region to be
about 40$^{\prime\prime}$ (height) $\times$ 40$^{\prime\prime}$ (length)
$\times$ 20$^{\prime\prime}$ (width) ($\sim 1.2 \times
10^{28}$~cm$^{3}$), and obtain the minimum density of about $1.4 \times
10^{10}$~cm$^{-3}$, which is not so small.
Therefore, the number density of about $10^{9}$~cm$^{-3}$ seems too
small based on the observations, although we cannot discard the
possibility of such small ambient plasma density at above-the-loop top
region where the HXR emission source appeared.

The actual HXR spectra may break at about 30~keV and steepen at the
higher energy range, which \citet{Hol03} reported in their spatially
integrated analysis, although it is unclear in our result.
(A HXR detector with a greater collecting area than that of RHESSI's
so that useful images can be obtained to higher energies and narrower
energy bands can be used would be a highly desirable feature for a
future instrument to make clear the issue.)
If such broken spectra are entirely applicable to the coronal HXR
emission source, it could be explained by the thick-thin model with
the electron spectral index $\delta$ of about 6.5, and the
thick-target should be responsible for the observed HXR emission.
Moreover, in that case, we may explain the gap of the spectral index
derived from the thick-thin HXR emission ($\delta_{H} \sim 6.5$) and
that from the microwave ($\delta_{\mu} \sim 4.7$) at higher energy
ranges, by considering that we may underestimate the thermal component
in microwaves.
Especially, the microwave emission at 34~GHz may suffer more from
the thermal contribution from cooler plasmas than detected by {\it
GOES}.
This gives a steeper microwave spectra than that estimated in this
paper.
To settle the uncertainty of the microwave spectral index, a microwave
interferometer with much high spectral resolution up to 40~GHz or more
is required.
However, even though, it seems that such a large electron spectral
index $\delta_{\mu}$ of about 6.5 is unexpected in the present case.

We finally have to note the difference of the positions of the microwave
and HXR emission sources.
The HXR emission source is located slightly higher than the microwave
emission source, as is often reported, and therefore, it is possible
that the HXR-emitting electrons are belong to the different group from
that for the microwave-emitting electrons.
The displacement between the microwave and HXR coronal emission
sources has been known for many flares, and the magnetic field
strength at the source position has been discussed to explain this.
Loop-top HXR sources are probably located above the SXR flare loops,
and therefore, the magnetic field there is weaker than the top of
closed flare loops.
Statistical studies on the displacement will be discussed in our
future papers.

\acknowledgments

We first acknowledge an anonymous referee for her/his useful comments 
and suggestions.
We wish to thank Drs. M. R. Kundu, E. J. Schmahl for fruitful
discussions and their helpful comments.
This work was carried out by the joint research program of the
Solar-Terrestrial Environment Laboratory, Nagoya University.
We made extensive use of {\it TRACE} and {\it RHESSI} Data Center.

\clearpage

\begin{table}
\begin{center}
\caption{Parameter survey for microwave emission source \label{tbl-1}}
\begin{tabular}{llll}\tableline\tableline
$B_{0}$ & $K$ constant$^{a}$ & $\theta$ & $N_{20keV}$ \\
gauss & ~ & deg & electrons cm$^{-3}$ \\
\tableline
100 & $7.8\times10^{14}$ & 60.7 & $3.1\times10^{9}$ \\
150 & $1.2\times10^{14}$ & 71.0 & $4.8\times10^{8}$ \\
200 & $3.6\times10^{13}$ & 75.8 & $1.4\times10^{8}$ \\
\tableline
\end{tabular}
\tablenotetext{a}{$dn_{\mu}(E)/dE = K E^{-\delta_{\mu}}$ electrons cm$^{-3}$}
\end{center}
\end{table}

\clearpage

\clearpage

\end{document}